\documentclass[12pt,a4paper]{article}
\usepackage{amssymb}
\usepackage{amsmath}
\usepackage{amsfonts,latexsym}
\usepackage{bbm}

\newcommand{\id}{\mathbbm{1}}

\font\mybb=msbm10 at 12pt
\def\bb#1{\hbox{\mybb#1}}
\font\mybbs=msbm10 at 10pt
\def\bbs#1{\hbox{\mybbs#1}}

\begin{document}

\rightline{UG-02/38}
\thispagestyle{empty}
\vskip 1truecm

\centerline{\LARGE \bf Type IIB Seven--brane Solutions}
\vskip .7truecm

\centerline{\LARGE \bf from}
\vskip .7truecm

\centerline{\LARGE \bf Nine-dimensional Domain Walls}
\vskip .7truecm

\centerline{\bf E.~Bergshoeff, U.~Gran and D.~Roest}
\bigskip
\centerline{Centre for Theoretical Physics, University of Groningen}
\centerline{Nijenborgh 4, 9747 AG Groningen, The Netherlands}
\centerline{E-mail: {\tt e.a.bergshoeff@phys.rug.nl, u.gran@phys.rug.nl,
d.roest@phys.rug.nl}}
\vskip 2truecm

\centerline{ABSTRACT}
\bigskip

We investigate half--supersymmetric domain wall solutions of four
maximally supersymmetric D=9 massive supergravity theories obtained
by Scherk--Schwarz reduction of D=10 IIA and IIB supergravity. One
of the theories does not have a superpotential and does not allow
domain wall solutions preserving any supersymmetry. The other
three theories have superpotentials
leading to half--supersymmetric domain wall solutions, one of which
has zero potential but non--zero superpotential.

The uplifting of these domain wall solutions to ten dimensions
leads to three classes of half--supersymmetric type IIB 7--brane solutions.
All solutions within each class are related by
$SL(2,\bb{R})$ transformations. The three classes together contain
solutions carrying all possible (quantized) 7--brane charges.
One class contains the
well-known D7--brane solution and its dual partners and we provide
the explicit solutions for the other two classes. The domain wall solution
with zero potential lifts up to a half--supersymmetric conical space--time.

\vfill\eject

\section{Introduction}

Recently, much attention has been given to the study of domain
wall solutions in (matter-coupled) supergravity theories. This is
due to several reasons. First of all, the possibility of a
supersymmetric RS scenario \cite{Randall:1999vf,Randall:1999ee}
relies on the existence of a special domain wall solution
containing a warp factor with the correct asymptotic behaviour
such that gravity is suppressed in the transverse direction.
Secondly, domain wall solutions play an important role in the
AdS/CFT correspondence \cite{Maldacena:1998re, Boonstra:1998mp}. 
A domain wall in D
dimensions may describe the renormalization group flow of the
corresponding field theory in D--1 dimensions. The geometrical
warp factor now plays the role of an energy scale. Finally, domain
wall solutions have been applied to cosmology, for some recent
papers see e.g.~\cite{Kallosh:2001gr,Townsend:2001ea}. In all
these cases the properties of the domain wall crucially depend on
the detailed properties of the scalar potential.

The highest-dimensional supergravity theory that allows a domain
wall solution is the maximally supersymmetric D=10 massive IIA
supergravity \cite{Romans:1986tz}. This theory is a massive
deformation, characterized by a mass parameter $m_{\rm R}$, of the
massless IIA supergravity theory
\cite{Campbell:1984zc,Giani:1984wc}. The particular domain wall
solution, the D8-brane, has been constructed in
\cite{Polchinski:1996df,Bergshoeff:1996ui}. In a supersymmetric
theory both the scalar potential $V$ as well as the massive
deformations in the supersymmetry transformations are often
characterized by a superpotential $W$. In the IIA case the
superpotential depends on just one scalar $\hat \phi$, the
dilaton, and is of a simple exponential form:
\begin{equation}
W(\hat\phi) = \tfrac{1}{4} e^{5\hat{\phi}/4} m_{\rm R} \, .
\label{sp10}
\end{equation}
In general, the lower-dimensional supergravity theories contain
more scalars and have correspondingly a more complicated
superpotential which is difficult to investigate. In fact, the
most general form of the superpotential is not always known
explicitly. In view of this, it is instructive to consider
maximally supersymmetric D=9 massive supergravity theories. These
theories on one hand share some of the complications of the
lower-dimensional supergravity theories and on the other hand are
simple enough to study in full detail.

The most general
Scherk-Schwarz reduction \cite{Scherk:1979ta} 
of D=10 IIB supergravity has been
considered in \cite{Meessen:1998qm}\footnote{A Scherk-Schwarz reduction
leading to
two mass parameters has been considered in \cite{Lavrinenko:1998qa}.}.
It leads to
$SL(2,\bb{R})$--covariant\footnote{Since our analysis below is at
the classical level we will work with $SL(2,{\bbs{R}})$ instead of
$SL(2,\bbs{Z})$. It is only in Section 5 that we will consider the
breaking of $SL(2,\bbs{R})$ to  $SL(2,\bbs{Z})$ at the quantum
level.} D=9 massive supergravity theories with mass parameters
$m_1, m_2$ and $m_3$.
By $SL(2,\bb{R})$ transformations one can go
to different mass parameters but the quantity $m_1{}^2+m_2{}^2-m_3{}^2$
is always invariant.
One therefore has three different theories depending on
whether this quantity is positive, negative or zero corresponding
to the three different conjugacy classes of $SL(2,\bb{R})$ \cite{Hull:1998vy}.
The supersymmetry
transformations of these massive supergravities have been
calculated recently \cite{Gheerardyn:2001jj}. The theory contains
three scalars $(\phi, \chi, \varphi)$ and we find that the
superpotential is given by:
\begin{equation}
  W_{\rm IIB}(\phi, \chi, \varphi) = {\tfrac{1}{4}} e^{2\varphi/\sqrt{7}}
  \biggl( m_2 \sinh(\phi) + m_3 \cosh(\phi) + m_1 e^\phi\chi
- {\textstyle{1\over 2}}(m_2-m_3)e^\phi \chi^2\biggr )\, .
\label{sp9}
\end{equation}
The scalar potential is given in terms of this superpotential
via the expression that follows from the positive energy requirement
\cite{Townsend:1984iu}:
\begin{equation}
  V =4 \left( \gamma^{AB}\frac{\delta W}{\delta\Phi^A}
  \frac{\delta W}{\delta \Phi^B}
  - \frac{D-1}{D-2} W^2 \right) \, .
\end{equation}
with $D=9$ and $\Phi^A=(\phi, \chi, \varphi)$ in this case.
Here $\gamma^{AB}$ is the inverse of the metric
$\gamma_{AB}$ occurring in the kinetic scalar term
$-\gamma_{AB}\partial\Phi^A\partial\Phi^B$.

In the IIA case, the situation is more subtle. We find that there are
two possibilities. Either one performs an (ordinary) Kaluza-Klein
reduction of D=10 massive IIA supergravity. This leads to a D=9
massive supergravity theory which is covered by the above superpotential
for the following choice of mass parameters (for more details, see
the next Section):
\begin{equation}
  m_1 = 0\, ,\hskip 1truecm m_2 = m_3 = m_{\rm R} \, .
\end{equation}
This is the massive T-duality of \cite{Bergshoeff:1996ui}.
The other possibility is to  set $m_{\rm R} = 0$ and perform a
generalized Scherk-Schwarz reduction making use of the $SO(1,1)$
symmetry of the action \cite{Lavrinenko:1998qa}. Since the
$SO(1,1)$--symmetry is only valid for $m_{\rm R} = 0$, i.e.~the
massive Romans deformation breaks the $SO(1,1)$--symmetry, one
cannot perform both reductions at the same time. The
Scherk-Schwarz reduction leads to an $SO(1,1)$--covariant
D=9 massive supergravity containing a single mass parameter $m_4$.
It turns out that in this case the massive deformations
cannot be expressed in terms of a  superpotential.

In this letter we study the domain wall
solutions allowed by the $SL(2,\bb{R})$-- and $SO(1,1)$--covariant
D=9 massive supergravities\footnote{Domain wall solutions of one
of the $SL(2,\bbs{R})$--covariant theories have been discussed in
\cite{Cowdall:2000sq}. We will compare our results with those of
\cite{Cowdall:2000sq} in Section 4.}. We find that the
$SO(1,1)$--covariant theory has no superpotential and does not
allow domain wall solutions. The other three
$SL(2,\bb{R})$-covariant theories have superpotentials that do
allow half--supersymmetric domain wall solutions.

The uplifting of these domain walls to ten dimensions
leads to three classes of half--supersymmetric type IIB 7--brane solutions.
All solutions within each class are related by
$SL(2,\bb{R})$ transformations and are characterized by two
holomorphic functions. The two functions are restricted by the
consistency requirement of yielding equal monodromy for the scalars and the Killing
spinors. We have explicitly checked that the solutions of our three classes
satisfy this requirement. These solutions give rise to all possible
 7--brane
charges. One class contains the well-known D7--brane solution and its
$SL(2,\bb{R})$--related partners.  We provide the previously unknown explicit
solutions for the two other classes. For each class we show which
solutions survive the quantization of $SL(2,\bb{R})$ to
$SL(2,\bb{Z})$.

We  find a special domain wall solution, corresponding to a zero
potential but non--zero superpotential. This half--supersymmetric domain
wall uplifts to either a fully supersymmetric Minkowski space-time or to 
half--supersymmetric conical type IIB solutions with deficit
angle $3 \pi/2$ or $5 \pi/3$ and without scalars. The conical
solutions have non-trivial monodromy due to the fermionic sector.

This paper is organized as follows. In Section 2 we review the IIA
theory in 10 and 9 dimensions and give the domain wall, or
D8-brane, solution of D=10 massive IIA supergravity. In Section 3
we discuss the IIB theory in 10 and 9 dimensions and give the
class of half--supersymmetric 7--brane solutions with two
holomorphic functions. In Section 4 we discuss the three classes
of D=9 domain wall solutions and their uplifting to ten
dimensions. The quantization conditions on the charges of the 7-branes
and mass parameters of the domain walls are discussed in Section 5. We
will summarize and discuss our results in the Conclusions. Our
conventions are given in Appendix A.

\section{IIA Supergravity in 10 and 9 Dimensions}

\subsection{D=10 Massive IIA Supergravity}

We first consider D=10 massive IIA supergravity. This theory
contains one scalar, the dilaton $\hat{\phi}$. For our purposes,
it is enough to consider only the kinetic terms for the graviton,
dilaton and R-R vector plus the mass term. In Einstein frame this
part of the Lagrangian reads:
\begin{align}
  \mathcal{L}_{\text{mIIA}}
  & = \tfrac{1}{2} \sqrt{-\hat{g}} \,
    \bigl[ -\hat{R} -\tfrac{1}{2} (\partial \hat{\phi})^2
    -\tfrac{1}{4} e^{3 \hat{\phi}/2} (\partial \hat A)^2 - V \bigr]
\label{mIIA}
\end{align}
with the potential $V$ given by a superpotential:
\begin{align}
  V = 8 (\tfrac{\delta W}{\delta \hat{\phi}})^2 - \tfrac{9}{2} W^2
    = \tfrac{1}{2} e^{5\hat{\phi}/2} m_{\rm R}^2 \text{~~~~ with~~}
  W = \tfrac{1}{4} e^{5\hat{\phi}/4} m_{\rm R} \, .
\end{align}
The corresponding supersymmetry transformations of the fermions are
\begin{equation}
\begin{aligned}
  \delta \hat{\psi}_{\hat{\mu}}
  & = \left (D_{\hat{\mu}} +\tfrac{1}{64} e^{3 \hat{\phi}/4} (\partial \hat A
)^{\hat{\nu} \hat{\rho}}
    (\hat{\Gamma}_{\hat{\mu} \hat{\nu} \hat{\rho}}
    - 14 \hat{g}_{\hat{\mu} \hat{\nu}} \hat{\Gamma}_{\hat{\rho}})
    \Gamma_{11} - \tfrac{1}{8} W \hat{\Gamma}_{\hat{\mu}}\right )
    \hat{\epsilon} \,, \\
  \delta \hat{\lambda} & = \left (
\not \! \partial \hat{\phi} +  \tfrac{3}{8} e^{3 \hat{\phi}/4}
    (\partial \hat A)^{\hat{\mu} \hat{\nu}} \hat{\Gamma}_{\hat{\mu}
\hat{\nu}} \Gamma_{11}
    + 4 \tfrac{\delta W}{\delta \hat{\phi}}\right )
    \hat{\epsilon} \,,
\end{aligned}
\label{susy10DmIIA}
\end{equation}
where $D_{\hat{\mu}}\hat\epsilon= (\partial_{\hat{\mu}}+
\hat{\omega}_{\hat \mu}) \hat\epsilon$ with the spin connection
$\hat{\omega}_{\hat \mu} = {\textstyle{1\over 4}}
\hat{\omega}_{\hat{\mu}}{}^{{\hat{a}} {\hat{b}}}
{\Gamma}_{{\hat{a}}{\hat{b}}}$. All spinors
$\hat{\psi}_{\hat{\mu}}, \hat{\lambda}, \hat{\epsilon}$ are real
Majorana spinors. The above transformation rules are the
Einstein-frame version of \cite{Bergshoeff:2001pv} and coincide
with those of \cite{Romans:1986tz} up to rescalings. Note that all
$m_{\rm R}$-dependent terms in both the Lagrangian and
transformation rules can be expressed in terms of the
superpotential \eqref{sp10}.

For $m_{\rm R}=0$ the Lagrangian is invariant under
$SO(1,1)$--transformations with weights as given in Table 1 (we
include all IIA fields and use Einstein frame metric). For $m_{\rm
R} \neq 0$ the Lagrangian is invariant if one also scales the mass
parameter $m_{\rm R}$ as indicated in Table 1.

\begin{table}[h]
\begin{center}
\begin{tabular}{||c|c|c|c|c|c|c|c|c||}
\hline \rule[-1mm]{0mm}{6mm}
Field       & $\hat{g}_{\hat\mu\hat\nu}$    & $\hat{B}_{\hat\mu\hat\nu}$ &
$e^{\hat{\phi}}$ & $\hat A_{\hat\mu}$ &
$\hat{C}_{\hat\mu\hat\nu\hat\rho}$ & $\hat{\psi}_{\hat\mu}$ & $\hat{\lambda}$ & $m_{\rm R}$   \\
\hline \hline \rule[-1mm]{0mm}{6mm}
$SO(1,1)$      & 0   &
${\textstyle{{1\over 2}}}$  &
1  &
$-{\textstyle{{3\over 4}}}$   &
$-{\textstyle{{1\over 4}}}$   &
$0$   &
$0$   &
$-{\textstyle{{5\over 4}}}$   \\
\hline
\end{tabular}
\caption{\it The $SO(1,1)$--weights of the IIA supergravity fields.}
\end{center}
\end{table}

The massive IIA supergravity theory has the D8-brane solution
\cite{Polchinski:1996df,Bergshoeff:1996ui}
\begin{align}
  \text{D8:~~~~~~}
  \hat{d}s^2  = H^{1/8} ds_9{}^2 + H^{9/8} dy^2 \,, \qquad
  e^{\hat{\phi}}  = H^{-5/4} \,, \qquad
  \text{with~~} H = 1+m_{\rm R} y\, .
\label{D8}
\end{align}
where we only consider $y$ such that $1+m_{\rm R} y$ is strictly positive
in order to have a well-behaved metric. By patching this solution at e.g. $y=0$
with an other solution having $H = 1-m_{\rm R} y$ a two-sided domain wall
positioned at $y=0$
can be obtained. In this letter we will always restrict to one side.
The D8-brane solution has the following non-zero spin connections
$(\hat{\mu} = (\mu,y))$ (for our conventions of underlined indices, see the
Appendix):
\begin{equation}
  \hat{\omega}_\mu =\tfrac{1}{32} H^{-25/16} \hat{\Gamma}_{\mu \underline{y}} m_{\rm R} \,, \qquad
  \hat{\omega}_y =0\, .
\end{equation}
It satisfies the Killing spinor equations \eqref{susy10DmIIA} for
\begin{align}
  (1-\Gamma_{\underline{y}}) \hat{\epsilon} = 0 \text{~~~~ with ~~}
  \hat{\epsilon} = H^{1/32} \hat{\epsilon}_0 \,,\hskip .5truecm
\hat{\epsilon}_0\ \ {\rm constant}\, .
\end{align}
Thus the D8-brane solution describes a 1/2 BPS state.

\subsection{IIA Reduction to 9 Dimensions}

We first consider the ordinary Kaluza-Klein reduction of
the massive IIA theory, i.e.~$m_{\rm R}\ne 0$, to 9 dimensions.
We use the following reduction rules for the bosons
(with $a=\tfrac{1}{8}$,
$b=-\tfrac{3}{8 \sqrt{7}}$, $c=\tfrac{3}{4}$, $d=\tfrac{\sqrt{7}}{4}$):
\begin{equation}
\begin{aligned}
  \hat{g}_{\mu\nu} & = e^{a \phi + b \varphi} g_{\mu\nu}\, , \\
  \hat{g}_{xx} & = e^{-7(a \phi + b \varphi)}\, , \qquad
\end{aligned} \qquad
\begin{aligned}
  \hat{\phi} & = c \phi + d \varphi\, ,  \\
  \hat{A}_x & = -2 \chi \,.
\end{aligned}
\end{equation}
The reduction rules of the fermions are given by
\begin{equation}
\begin{aligned}
  \hat{\psi}_\mu & = e^{(a\phi+b\varphi)/4}
    (\psi_\mu + \tfrac{1}{4} \Gamma_{\mu} (a \lambda + b \tilde{\lambda})) \,, \\
  \hat{\psi}_{\underline{x}} & = - \tfrac{7}{4} \Gamma_{\underline{x}}
    e^{-(a\phi+b\varphi)/4} (a \lambda + b \tilde{\lambda}) \,, \qquad
\end{aligned} \qquad
\begin{aligned}
  \hat{\lambda} & = e^{-(a\phi+b\varphi)/4} (c \lambda + d \tilde{\lambda})\,, \\
  \hat{\epsilon} & = e^{(a\phi+b\varphi)/4} \epsilon \,.
\end{aligned}
\label{redfermIIA}
\end{equation}
The scalar dependence is put in such that we obtain the
conventional form of the D=9 supersymmetry rules corresponding to
a standard kinetic term for the D=9 graviton and gravitini. This
also explains the mixing between $\psi_\mu$ and $\tilde{\lambda}$
in the first line, which implies that only $\delta
\tilde{\lambda}$ contains $\partial \varphi$ terms. Thus the
massive IIA theory \eqref{mIIA} reduces to the D=9 massive
Lagrangian
\begin{align}
  \mathcal{L}
  & = \tfrac{1}{2} \sqrt{-{g}} \,
    [ -{R} - \tfrac{1}{2} (\partial \phi)^2 - \tfrac{1}{2} (\partial \varphi)^2
    - \tfrac{1}{2} e^{2 \phi} (\partial \chi)^2
    - V]
\end{align}
with the potential $V$ given by a superpotential:
\begin{align}
  V = 8 (\tfrac{\delta W}{\delta \phi})^2 + 8 (\tfrac{\delta W}{\delta \varphi})^2
       - \tfrac{32}{7} W^2
    = \tfrac{1}{2} e^{2 \phi + 4 \varphi/\sqrt{7}} m_{\rm R}^2 \text{~~~~ with~~}
  W = \tfrac{1}{4} e^{\phi + 2 \varphi/\sqrt{7}} m_{\rm R} \,,
\end{align}
The supersymmetry rules \eqref{susy10DmIIA} reduce to
\begin{equation}
\begin{aligned}
  \delta \psi_\mu & = (D_\mu
    +\tfrac{1}{4} e^\phi \partial_\mu \chi \Gamma_{\underline{x}} \Gamma_{11}
    -\tfrac{1}{7} W \Gamma_\mu ) \epsilon \,,  \\
  \delta \lambda
  & = (\not \! \partial \phi\,  - e^\phi \! \! \not \! \partial \chi \Gamma_{\underline{x}} \Gamma_{11}
    + 4 \tfrac{\delta W}{\delta \phi} ) \epsilon \,, \\
  \delta \tilde{\lambda}
  & =  (\not \! \partial \varphi + 4 \tfrac{\delta W}{\delta \varphi} ) \epsilon \,.
\end{aligned}
\end{equation}
Note that the massive deformations of both the Lagrangian and
supersymmetry rules can be given in terms of a superpotential, as
in 10 dimensions. Later, in Section 3, we will see that the above
Lagrangian and transformation rules can also be obtained via a
particular Scherk--Schwarz reduction of D=10 IIB supergravity.

There is another massive 9D theory that can be obtained from reducing IIA
supergravity. To obtain this 9D theory
 one has to use the $SO(1,1)$ scale symmetry of the 10D theory
 \cite{Lavrinenko:1998qa}.
This symmetry
implies the consistency of the generalized reduction rules with a specific
$x$-dependence of the 10D fields, depending on their $SO(1,1)$ weights.
This introduces a new mass parameter, which we call $m_4$,
 upon reduction to 9 dimensions.
Since the $SO(1,1)$ symmetry is broken by nonzero $m_{\rm R}$
(unless one scales it), the generalized reduction is only applicable to
 massless 10D IIA supergravity.
The generalized reduction rules read
\begin{equation}
\begin{aligned}
  \hat{g}_{\mu\nu} & = e^{a \phi + b \varphi} g_{\mu\nu}\, , \\
  \hat{g}_{xx} & = e^{-7(a \phi + b \varphi)}\, , \qquad
\end{aligned} \qquad
\begin{aligned}
  \hat{\phi} & = c \phi + d \varphi + m_4 x\, , \\
  \hat{A}_x & = - 2 e^{-3m_4 x/4} \chi \,,
\end{aligned}
\end{equation}
with $a,b,c,d$ as above  and with the fermionic reduction rules as
in the previous case, independent of $x$. Thus we find the
following $SO(1,1)$--covariant D=9 massive
Lagrangian\footnote{Strictly speaking, the Lagrangian below is
only $SO(1,1)$--invariant for $m_4=0$. To obtain manifest
$SO(1,1)$--invariance one should replace $m_4$ by a scalar field
via a Lagrange multiplier, see the Conclusions. The same remark
applies to the $SL(2,\bbs{R})$--covariant massive supergravity
theory of Section 4.}:
\begin{align}
  \mathcal{L}
  & = \tfrac{1}{2} \sqrt{-{g}} \,
   \left[ -{R} - \tfrac{1}{2} (\partial \phi)^2 - \tfrac{1}{2} (\partial \varphi)^2
    - \tfrac{1}{2} e^{2 \phi} (\partial \chi)^2
    - \tfrac{1}{2} e^{\phi -3 \varphi / \sqrt{7} } m_4^2\right]
\end{align}
with the supersymmetry rules
\begin{equation}
\begin{aligned}
  \delta \psi_\mu & = (D_\mu
    + \tfrac{1}{4} e^\phi \partial_\mu \chi \Gamma_{\underline{x}} \Gamma_{11}
    ) \epsilon \,, \\
  \delta \lambda
  & = (\not \! \partial \phi\, - e^\phi \not \! \partial \chi \Gamma_{\underline{x}} \Gamma_{11}
    +\tfrac{3}{4} m_4 e^{\phi/2 -3 \varphi /2 \sqrt{7}} \Gamma_{\underline{x}}) \epsilon \,, \\\
  \delta \tilde{\lambda}
  & =  (\not \! \partial \varphi
    +\tfrac{\sqrt{7}}{4} m_4 e^{\phi/2 -3 \varphi /2 \sqrt{7}} \Gamma_{\underline{x}}) \epsilon \,.
\end{aligned}
\label{class4susy}
\end{equation}
A peculiar feature of this massive 9D theory is that the potential does not have a corresponding superpotential.

\section{IIB Supergravity in 10 and 9 Dimensions}

\subsection{D=10 IIB Supergravity}

We next consider  D=10 IIB supergravity. This theory has two
scalars, a dilaton $\hat\phi$ and an axion $\hat\chi$. We truncate
to the gravity-scalar part. This part of the Lagrangian reads in
Einstein frame:
\begin{equation}
\begin{aligned}
  \mathcal{L}_{\text{IIB}}
  & = \tfrac{1}{2} \sqrt{-\hat{g}} \,
    [ -\hat{R} -\tfrac{1}{2} (\partial \hat{\phi})^2
    -\tfrac{1}{2} e^{2 \hat{\phi}} (\partial \hat{\chi})^2 ]  \\
  & = \tfrac{1}{2} \sqrt{-\hat{g}} \,
    [ -\hat{R} + \tfrac{1}{4} \text{Tr}
    (\partial \hat{\mathcal{M}} \partial \hat{\mathcal{M}}^{-1}) ] \,.
\end{aligned}
\label{10DIIB}
\end{equation}
The two scalars $\hat{\phi}$ and $\hat{\chi}$ parametrize a
$SL(2,\bb{R})/SO(2)$ coset space as follows:
\begin{equation}
  \hat{\mathcal{M}} = e^{\hat{\phi}} \left(
  \begin{array}{cc} |\hat{\tau}|^2 & \hat{\chi} \\ \hat{\chi} & 1 \end{array}
  \right) ~~~~ \text{with} ~~~ \hat{\tau} = \hat{\chi} + i e^{-\hat{\phi}} \,.
\end{equation}
The SL(2,\bb{R}) duality acts in the following way:
\begin{equation}
  \mathcal{\hat{M}} \rightarrow \Omega \, \mathcal{\hat{M}} \, \Omega^T \,, \qquad
  \text{or~~} \hat{\tau} \rightarrow \frac{a \hat{\tau} +b}{c\hat{\tau} +d} \,, \qquad
  \text{with~~} \Omega = \left( \begin{array}{cc} a&b\\c&d \end{array} \right) \in SL(2,\bb{R}) \,.
\label{Sduality}
\end{equation}
For later use we give the two elements whose products span $SL(2,\bb{Z})$:
\begin{align}
  S= \left( \begin{array}{cc} 0&1\\-1&0 \end{array} \right) \,, \qquad
  T= \left( \begin{array}{cc} 1&1\\0&1 \end{array} \right) \,.
\label{SandT}
\end{align}
The Einstein frame metric is $SL(2,\bb{R})$--invariant.
The corresponding truncated supersymmetry variations of the fermions read
\begin{equation}
\begin{aligned}
  \delta \hat{\psi}_{\hat{\mu}}
  & = (D_{\hat{\mu}} + \tfrac{1}{4} i e^{\hat{\phi}}
    \partial_{\hat{\mu}} \hat{\chi} )\, \hat{\epsilon} \,,  \\
  \delta \hat{\lambda} & = (\not \! \partial \hat{\phi}
  + ie^{\hat{\phi}} \! \not \! \partial \hat{\chi} ) \, \hat{\epsilon}^* \,,
\end{aligned}
\label{10DIIBsusy}
\end{equation}
where $D_{\hat{\mu}}\hat\epsilon =
(\partial_{\hat{\mu}}+ \hat{\omega}_{\hat \mu})\hat\epsilon$ with
$\hat{\omega}_{\hat \mu} = {\textstyle{1\over 4}} \hat{\omega}_{\hat{\mu}}{}^{{\hat{a}}
{\hat{b}}} {\Gamma}_{{\hat{a}}{\hat{b}}}$
the spin connection. All spinors
$\hat{\psi}_{\hat{\mu}}, \hat{\lambda}, \hat{\epsilon}$ are
complex Weyl spinors. The fermions transform under the SL(2,\bb{R})
transformation \eqref{Sduality} as \cite{Bergshoeff:1996ui}\footnote{Note that
the duality transformations of both the scalars and the fermions do not
change if we replace $\Omega$ by $-\Omega$. Therefore these
fields transform under $PSL(2,\bb{R})$. From now on we will only
consider group elements $\Omega$ that are continuously connected
to the unit element.}
\begin{align}
  \hat{\psi}_{\hat{\mu}} & \rightarrow \left( \frac{c \, \hat{\tau}^*+d}{c
\, \hat{\tau}+d} \right)^{1/4}
    \hat{\psi}_{\hat{\mu}} \,, \qquad
  \hat{\lambda} \rightarrow \left( \frac{c \, \hat{\tau}^*+d}{c \, \hat{\tau}+d}
\right)^{3/4}
    \hat{\lambda} \,, \qquad
  \hat{\epsilon} \rightarrow \left( \frac{c \, \hat{\tau}^*+d}{c \, \hat{\tau}+d}
 \right)^{1/4}
    \hat{\epsilon} \,.
\label{fermionSduality}
\end{align}
In particular, they are invariant under the shift symmetry
$\hat\chi \rightarrow \hat\chi + b$ which has $a=d=1$ and $c=0$
and the scale symmetry $\hat \tau \rightarrow a^2 \hat \tau$ which has
$d=a^{-1}$ and $b=c=0$.

\subsection{Half--supersymmetric 7--brane Solutions}

The D=10 IIB supergravity theory allows for a family of
1/2--supersymmetric seven--brane solutions containing two
functions $f$ and $g$, which are seperately (anti-)holomorphic.
For notational clarity we will always take both $f$ and $g$ to be
holomorphic. In these solutions the scalar $\hat{\tau}$ is given
by the function $f$ \cite{Greene:1990ya,Gibbons:1996vg}. This
function determines the monodromy of the scalars. The second
function $g$ appears only in the metric and in the Killing spinor.
The monodromy of the Killing spinor is determined by $f$ and $g$.
The requirement that the monodromies of the scalars and the
Killing spinor coincide puts restrictions on $f$ and $g$. The
function $g$ can always be transformed away by a holomorphic
coordinate transformation\footnote{This and related issues have
been discussed independently by Tom\'as Ort\'\i n in unpublished
notes.} and only affects global issues like monodromy and deficit
angle. The occurance of this function in the metric was already
considered in \cite{Greene:1990ya, Meessen:1998qm}. The general
solution with two holomorphic functions reads\footnote{The solutions 
we consider generically do not have finite energy. 
To obtain a globally well-defined, finite-energy,
solution one should use the so-called $j(\tau)$-function as explained in
\cite{Greene:1990ya}.}

\begin{align}
  \hat{ds}^2 & = ds_8{}^2 + \text{Im}(f)\,e^{-\text{Re}(g)} dz d\overline{z} \,,
  \hskip 2truecm \hat \tau  =  f\, ,
\label{7brane}
\end{align}
with the holomorphicity conditions $\partial_{\overline z} f =
\partial_{\overline z} g=0$. The general seven-brane solution \eqref{7brane}
has the spin connection $(\hat \mu = (\mu,z,\overline{z}))$:
\begin{equation}
  \hat{\omega}_\mu=0 \,, \qquad
  \hat{\omega}_i = \tfrac{1}{2} \Gamma_{\underline{ij}}
  \left(
  \text{Im} (f)^{-1} \partial_i \text{Im} (f)
  - \partial_i \text{Re}(g)
  \right)
  \qquad i,j=(z,\overline{z}) \, .
\end{equation}
The solution \eqref{7brane} satisfies the Killing spinor equations \eqref{10DIIBsusy} for
\begin{align}
  \Gamma_{\underline{z}} \hat{\epsilon} = 0 \text{~~~~ with ~~}
    \hat{\epsilon} = e^{i \text{Im}(g)/4} \hat{\epsilon}_0 \,.
\label{Killing7brane}
\end{align}
Thus the general 7-brane solution preserves 1/2 of supersymmetry.
The special case that $\partial_i \text{Im}(f) = 0$ (thus
implying that $f$ is constant) can lead to an
enhancement of supersymmetry. This case will be treated at the end
of this Section.

The holomorphic function $g(z)$ can be eliminated locally from the
general 7--brane solution \eqref{7brane} via the holomorphic coordinate
transformation
\begin{align}
  z' = \int_{z_0}^{z_0+z} d \tilde{z} \, e^{-\xi(\tilde{z})/2 }\,
\label{holomorphicgct}
\end{align}
for $\xi(z) = g(z)$.
This also transforms the Killing spinor \eqref{Killing7brane}
to a space-time independent constant spinor $\hat \epsilon_0$.
More generally, given a 7--brane solution
with functions $f(z), g(z)$ the holomorphic coordinate transformation
\eqref{holomorphicgct} gives us an equivalent 7--brane solution with
$f^\prime(z^\prime) = f(z)$ and $g^\prime (z') = g(z) - \xi(z)$.
Note that although the holomorphic function $g$ can be transformed away
locally it may have implications on global issues like the monodromy
and the deficit angle. We note that the choice $g=-2 B_1 \log(z)$
leads to the 7--brane solutions of \cite{Einhorn:2000ct}.

Under an $SL(2,\bb{R})$--transformation a
7--brane solution \eqref{7brane}
is transformed into another member of the same class \eqref{7brane}.
In particular, under the $SL(2,\bb{R})$--transformations \eqref{Sduality}
the holomorphic functions transform as
\begin{align}
  f \rightarrow \frac{af+b}{cf+d} \,, \qquad
  g \rightarrow g - 2 \log(cf+d) \,.
\end{align}
This relates for example
the D7-brane solution to its S-dual partner, the Q7-brane
\cite{Meessen:1998qm} via an S--transformation \eqref{SandT}:
\begin{align}
  \text{D7:~~} \begin{array}{l} f=i\, m \log(-iz) \\ g=0 \end{array}
  \overset{\textstyle S}{\longrightarrow} \;\;
  \text{Q7:~~} \begin{array}{l} f=(-i\, m \log(-iz))^{-1} \\
    g=-2 \log(-i\, m \log(-iz)) \end{array} \,.
\label{conD7toQ7}
\end{align}

It is conventional to use polar coordinates for 7-branes:
$z=re^{i\theta}$ with $0<r<\infty$ and $0\leq \theta< 2\pi$. The
D7-brane given above is an example of this. The 7-brane is located
at $r=0$ and therefore the monodromy is determined by going round
in the $\theta$ direction~\cite{Meessen:1998qm}. The deficit angle
can be determined by going to Minkowski space-time
locally~\cite{Dabholkar:1997zd}. For the purpose of dimensional
reduction we find it more convenient to use cylindrical
coordinates: $z=x+iy$ with $x \simeq x + 2\pi R$.
The monodromy is then determined by the relation
between the fields at $x$ and at $x+ 2 \pi R$. The cylindrical
coordinates $z$ are related to the polar coordinate $z'$ by the
holomorphic coordinate transformation \eqref{holomorphicgct} where
$\xi(z) = 2 i z/R + 2 \log(R)$. Thus the cylindrical D7-brane scalars
$f'(z') =i\, m' \log(-iz')$ read in polar coordinates $f(z) = mz$ with
$m=m'/R$ \cite{Bergshoeff:1996ui}.
From now on we will use cylindrical coordinates unless explicitly
indicated otherwise.

The $SL(2,\bb{R})$ monodromy of the scalars and the Killing spinors
corresponding to the general solution \eqref{7brane} can be inferred from
the relation between the fields at $x$ and $x+ 2\pi R$. From the
transformations \eqref{Sduality} and \eqref{fermionSduality}
we can read off the relations
\begin{align}
  \hat \tau (x+2\pi R) &
    = \frac{a \hat \tau (x)+b}{c \hat \tau(x)+d} \,, \qquad
  \hat \epsilon (x+2 \pi R)
    = \left( \frac{c \hat \tau (x)^* +d}{c \hat \tau(x)+d} \right)^{1/4} \,
    \hat \epsilon (x) \,.
\label{monodromy}
\end{align}
It is convenient to parametrise the monodromy matrix $\Lambda$ by
\begin{align}
  \Lambda &
    = \left(
    \begin{array}{cc} a & b \\ c & d \end{array}
    \right)
    = e^{2 \pi R \, C}
     ~~~ \text{with} ~~~ C = \tfrac{1}{2} \left(
    \begin{array}{cc} m_1 & m_2+m_3 \\ m_2-m_3 & -m_1 \end{array}
    \right) \, ,
\label{charges}
\end{align}
where $2 \pi R \, C$ is a linear combination of the three generators
of $SL(2,\bb{R})$. The constants $\vec{m} = (m_1,m_2,m_3)$ can be seen as the
different charges of the 7-brane solution in some basis \cite{Meessen:1998qm}.
These charges are determined by the monodromy of the function $f(z)$.
For example the cylindrical D7-brane with  $f(z) = mz$ leads to the
monodromy relations
\begin{align}
  f(z+2\pi R) = f(z) + 2\pi mR \; \; \Rightarrow \;
  \Lambda = \left( \begin{array}{cc} 1 & 2\pi mR \\ 0 & 1\end{array} \right) \; \; \Rightarrow \;
  \vec{m} = (0,m,m) \,.
\end{align}
using \eqref{Sduality} and \eqref{charges}.

Acting with an $SL(2,\bb{R})$--transformation \eqref{Sduality}
on the scalars amounts to the transformation of the monodromy matrix
\begin{align}
  \Lambda \rightarrow \Omega \, \Lambda \, \Omega^{-1} \,, \;\; \text{or~~}
  C \rightarrow \Omega \, C \, \Omega^{-1} \,.
\label{monodromytransf}
\end{align}
Note that this leaves $\alpha^2 = -\det(C) = {\textstyle
{1\over 4}} (m_1{}^2+m_2{}^2-m_3{}^2)$ invariant. Thus all
$SL(2,\bb{R})$ related 7--brane solutions have the same value of
$\alpha^2$. Thus for the D7-brane and for all other 7--branes
related to the D7--brane via an $SL(2,\bb{R})$--transformation we
find $\alpha^2=0$. In Section 4 we will see that the uplifting of
certain D=9 domain wall solutions will give us examples of
7--brane solutions with $\alpha^2$ positive and negative as well.

Let us finally comment on the case of constant scalars, i.e.~constant $f$.
The solution \eqref{7brane} then becomes purely gravitational
and has a second Killing spinor given by
\begin{align}
  \Gamma_{\overline{\underline{z}}} \hat{\epsilon} = 0 \text{~~~~ with ~~}
    \hat{\epsilon} = e^{-i\text{Im}(g)/4} \hat{\epsilon}_0 \,.
\end{align}
The two Killing spinors build up a full $N=2$ spinor. However, for
the gravitational solution with constant $f$ to have unbroken supersymmetry
one must require equal monodromies for the two Killing spinors.
The gravitational solution can be related locally to a Minkowski space-time via the
coordinate transformation \eqref{holomorphicgct} with $\xi(z)=g(z)$
but global issues may prevent the identification with Minkowski space-time.
This depends on the boundary conditions on $g$. We will see an explicit
example of this in Section 6 where we will encounter a half--supersymmetric 
conical space-time solution.

\subsection{IIB Reduction to 9 Dimensions}

We now derive the relevant part of the $SL(2,\bb{R})$-covariant
N=2, D=9 massive supergravity theories by performing a generalized
Scherk-Schwarz reduction of the truncated IIB supergravity
Lagrangian \eqref{10DIIB}. For more details, see
\cite{Meessen:1998qm,Gheerardyn:2001jj}. To be specific, we make
the following IIB reduction Ans\"atze $(\hat{\mu} = (\mu,x))$:
\begin{equation}
\begin{aligned}
  \hat{g}_{\mu\nu} & = e^{\sqrt{7} \varphi /14} g_{\mu\nu}\, , \\
  \hat{g}_{xx} & = e^{-\sqrt{7} \varphi /2}\, ,  \\
  \hat{\mathcal{M}} & = \Omega(x) \mathcal{M} \, \Omega(x)^T \,,
\end{aligned}
\label{redIIBto9}
\end{equation}
where we have given the D=10 dilaton $\hat\phi$ and axion
$\hat\chi$ an $x$-dependence via the
$SL(2,\bb{R})$-element\footnote{The precise rule for assigning the
$x$-dependence is: (i) replace $\Omega$ by $\Omega(x)$ in the D=10
$SL(2,\bbs{R})$--transformation rule and (ii) replace the D=10
fields occurring in the transformation rule by $x$--independent
D=9 fields.}
\begin{align}
\Omega (x)  & = e^{x C} = \left(
    \begin{array}{cc} \cosh(\alpha x) + \tfrac{m_1}{2\alpha} \sinh(\alpha x) &
     \ \ \ \tfrac{m_2+m_3}{2 \alpha} \sinh(\alpha x) \\
        &\\
     \tfrac{m_2-m_3}{2 \alpha} \sinh(\alpha x) &
     \ \ \ \cosh(\alpha x) - \tfrac{m_1}{2\alpha} \sinh(\alpha x) \end{array}
    \right)
\label{redtransf}
\end{align}
with $\alpha$ and $C$ defined in the previous subsection.
Note that this reduction Ansatz implies the identification
of the monodromy matrix of 7-brane solutions in 10D with the mass
matrix of domain walls in 9D.
Thus the charges of the 7-branes
provide the masses of the domain walls upon reduction \cite{Meessen:1998qm}.

These reduction Ans\"atze lead to the following truncated N=2, D=9
$SL(2,\bb{R})$-covariant massive supergravity Lagrangian\footnote{
Strictly speaking the D=9 Lagrangian is also covariant under an
additional $SO(1,1)$ which acts on the scalars as $\varphi^\prime
= \varphi + c$ for constant $c$. Therefore the full symmetry group
is $GL(2,\bbs{R}) = SL(2,\bbs{R}) \otimes SO(1,1)$.}:
\begin{align}
  \mathcal{L}_{\text{9D}}
    = \tfrac{1}{2} \sqrt{-{g}} \,
    \bigl [ -{R} +
  & \tfrac{1}{4} \text{Tr}
    (\partial {\mathcal{M}} \partial {\mathcal{M}}^{-1})
    -\tfrac{1}{2} (\partial \varphi)^2 - V(\phi, \chi, \varphi)\bigr ]\, .
\label{m9DIIBaction}
\end{align}
The potential $V(\phi, \chi, \varphi)$ is given by
\begin{equation}
\begin{aligned}
  V(\phi, \chi, \varphi)
  & = \tfrac{1}{2} e^{4 \varphi / \sqrt{7}} \text{Tr}
    ( C^2 +  C \mathcal{M}^{-1} C^T \mathcal{M} ) \,, \\
  & = 8 (\tfrac{\delta W}{\delta \phi})^2
    +8 \, e^{-2\phi} (\tfrac{\delta W}{\delta \chi})^2
    +8 (\tfrac{\delta W}{\delta \varphi})^2
    -{\textstyle{{32\over 7}}} W^2 \, ,
\end{aligned}
\label{pot9B}
\end{equation}
with the superpotential $W(\phi, \chi, \varphi)$:
\begin{equation}
    W(\phi, \chi, \varphi) = {\textstyle{1\over 4}} e^{2\varphi/\sqrt{7}}
    \left( m_2 \sinh(\phi) + m_3 \cosh(\phi) + m_1 e^\phi\chi
    - {\textstyle{1\over 2}}(m_2-m_3)e^\phi \chi^2\right )\, .
\label{sp9revisited}
\end{equation}
The supersymmetry transformations corresponding to the D=9
massive action \eqref{m9DIIBaction} follow
from reducing the massless D=10 supersymmetry rules \eqref{10DIIBsusy} with the reduction Ans\"atze
\begin{equation}
\begin{aligned}
  \hat{\psi}_\mu & = e^{\sqrt{7} \varphi /56}
    \left( \frac{c \tau^* + d}{c \tau + d} \right)^{1/4}
    (\psi_\mu + \tfrac{1}{8\sqrt{7}} \Gamma_{\mu} \tilde{\lambda}^* )\,, \\
  \hat{\psi}_{\underline{x}} & = - \tfrac{\sqrt{7}}{8} \Gamma_{\underline{x}}
    e^{-\sqrt{7} \varphi /56}
    \left( \frac{c \tau^* + d}{c \tau + d} \right)^{1/4} \tilde{\lambda}^* \,, \qquad
\end{aligned} \qquad
\begin{aligned}
  \hat{\lambda} & = e^{-\sqrt{7} \varphi /56}
    \left( \frac{c \tau^* + d}{c \tau + d} \right)^{3/4} \lambda \,,  \\
  \hat{\epsilon} & = e^{\sqrt{7} \varphi /56}
    \left( \frac{c \tau^* + d}{c \tau + d} \right)^{1/4} \epsilon \,.
\end{aligned}
\label{redf}
\end{equation}
We have given the D=10 fermions an $x$-dependence via the same
$SL(2,\bb{R})$ element \eqref{redtransf}, i.e.~the values of $c$ and $d$
in \eqref{redf} are given by
\begin{equation}
  c=\tfrac{m_2-m_3}{2 \alpha} \sinh(\alpha x) \,, \qquad
  d=\cosh(\alpha x) - \tfrac{m_1}{2 \alpha} \sinh(\alpha x) \,.
\end{equation}
The same considerations concerning the $e^{\varphi}$-dependence and
mixing of $\psi_\mu$ and $\tilde{\lambda}^*$ apply as in the IIA case
\eqref{redfermIIA}. The $x$-dependence via $c$ and $d$ is put in to
ensure that the 9D theory is independent of $x$.
With these reduction Ans\"atze we obtain the following D=9 massive
supersymmetry rules
\begin{equation}
\begin{aligned}
  \delta \psi_\mu & = (D_\mu
    +\tfrac{i}{4} e^\phi \partial_\mu \chi
    +\tfrac{i}{7} \Gamma_{\mu \underline x} W ) \epsilon \,, \\
  \delta \lambda
  & = (\not \! \partial \phi\,  + 4 i \Gamma_{\underline x} \tfrac{\delta W}{\delta \phi}
    + i e^\phi (\not \! \partial \chi + 4 i \Gamma_{\underline x} e^{-2 \phi}
    \tfrac{\delta W}{\delta \chi}) ) \epsilon^* \,, \\
  \delta \tilde{\lambda}
  & =  (\not \! \partial \varphi + 4 i \Gamma_{\underline x}
    \tfrac{\delta W}{\delta \varphi} ) \epsilon^*,
\end{aligned}
\label{D9susyrules}
\end{equation}
with the superpotential given by \eqref{sp9revisited}. These were also derived
by \cite{Gheerardyn:2001jj}.

The inclusion of the three mass parameters breaks the $SL(2,\bb{R})$
invariance. Rather the duality transformation now maps between
theories with different mass parameters:
\begin{align}
  C \rightarrow (\Omega^T)^{-1} \, C \, \Omega^T \,.
\end{align}
It is in this sense that the theory is covariant under $SL(2,\bb{R})$
transformations.
Note that the relation is of the same form as \eqref{monodromytransf}:
in fact the duality relations between 7--branes in 10D
and domain walls in 9D is identical. Again this transformation
preserves $\alpha^2 = {\textstyle {1\over 4}} (m_1{}^2+m_2{}^2-m_3{}^2)$.
Thus one must distinguish three different theories depending on
whether $\alpha^2$ is positive, negative or zero corresponding
to the three different conjugacy classes of $SL(2,\bb{R})$ \cite{Hull:1998vy}.
For each class it is convenient to make
a specific choice of basis for $\vec{m}=(m_1,m_2,m_3)$. For later use
we give the explicit form of the potential in each class:
\begin{equation}
\begin{aligned}
{\rm Class\ I}:\ \ \ &\alpha^2 = 0\, ,\ \   \vec{m}=(0,m,m): \\
  & V(\phi,\varphi,\chi) = \tfrac{1}{2} e^{4 \varphi / \sqrt{7} +
 2 \phi} m^2 \,, \\
&\\
{\rm Class\ II}:\ \ \ &\alpha^2 > 0\, ,\ \   \vec{m}=(m,0,0):\\
  & V(\phi,\varphi,\chi) = \tfrac{1}{2} e^{4 \varphi / \sqrt{7} }
    (1+e^{2 \phi} \chi^2) m^2 \,,
 \\
&\\
{\rm Class\ III}:\ \ \ &\alpha^2 < 0\, ,\ \   \vec{m}=(0,0,m):\\
  & V(\phi,\varphi,\chi) = \tfrac{1}{2} e^{4 \varphi / \sqrt{7}}
    \left (\sinh^2(\phi)+ \chi^2 (2+e^{2 \phi}(2+ \chi^2))\right) \, m^2 \,.
\end{aligned}
\label{3classes}
\end{equation}

Comparing with the IIA results one finds that for the values
$\vec{m}=(0,m_{\rm R},m_{\rm R})$ (class I) the reduction of IIB supergravity
equals the reduction of
massive IIA supergravity.
Also the superpotentials and hence the supersymmetry transformations
 are equal for these values of the mass parameters.
This corresponds to the massive T-duality between the
D8-brane solution \eqref{D8} and the D7-brane solution \cite{Bergshoeff:1996ui}.
The other massive deformation of IIA, coming from the $SO(1,1)$ scale
symmetry, can not be reproduced by the IIB reduction.
This is obvious from the lack of a superpotential at the IIA side.
Thus one can construct four different massive deformations of D=9,
N=2 supergravity from considering both its IIA and IIB origin.

\section{Domain Wall Solutions and their Upliftings}

We are now ready to investigate domain wall solutions for the
three classes of nine-dimensional massive supergravity theories
coming from the IIB side (class I-III) and the massive supergravity
theory coming from the IIA side (class IV). We do not consider
seperately the theory obtained by reducing 10D massive IIA
supergravity since, as mentioned above, this 9D theory coincides
with class I if we set $m_2=m_3=m_R$ and $m_1=0$.

We will start by constructing half-supersymmetric solutions
to the Killing spinor equations in 9 dimensions that follow from
the supersymmetry rules (\ref{D9susyrules}) and (\ref{class4susy}).
Our only input will be a domain wall Ansatz, i.e. we assume a
diagonal $8+1$ split of the metric with all fields depending only
on the single transverse $y$--direction. These solutions
automatically define half--supersymmetric domain wall solutions to
the full equations of motion. After that we will uplift these
solutions to 10 dimensions. We find that all 10D 7--branes fall in
the general class \eqref{7brane} with two holomorphic functions,
as indicated in Table 2 \footnote{
  For clarity we have taken the constants $C_1$ and $C_2$ which
appear later equal to $C_1=1$ and $C_2=0$.}. The D=9 domain walls correspond to the
potential \eqref{pot9B} with mass parameters
$\vec{m}=(m_1,m_2,m_3)$. These mass parameters automatically
define the charge of the D=10 7-brane solutions
\cite{Meessen:1998qm}.

\begin{table}[ht]
\begin{center}
\begin{tabular}{||c|c|c|c|c|c||}
\hline \rule[-1mm]{0mm}{6mm}
  Class & $\alpha^2$ & $\vec{m}$ & & $f(z)$ & $g(z)$ \\
\hline \hline \rule[-1mm]{0mm}{6mm}
  I & 0 & $(0,m,m)$ & D7: & $mz$ & $0$ \\
\hline \rule[-1mm]{0mm}{6mm}
  II & $\tfrac{1}{2} m^2$ & $(m,0,0)$ & R7: & $i e^{mz}$ & $m z$ \\
\hline \rule[-1mm]{0mm}{6mm}
  III & $-\tfrac{1}{2} m^2$ & $(0,0,m)$ & T7: & $\text{tan}(\tfrac{1}{2}mz)$
    & $-2\log\left(\cos({1\over 2}mz)\right)$ \\
  & & & G7: & $i$ & $i m z$ \\
\hline
\end{tabular}
\caption{\it The table indicates the different solutions for the three classes.
  It gives the $\vec{m}$ charges and the functions $f(z)$ and $g(z)$ of
  the D=10 7--brane solutions that follow from uplifting of the D=9
domain walls.}
\end{center}
\end{table}

We find that there are three independent 7--brane solutions
carried by scalars: D7, R7 and T7. These cannot be related by
$SL(2,\bb{R})$--transformations since their charges give rise to
different $SL(2,\bb{R})$--invariants $\alpha^2$. Unlike the
well-studied D7-brane solution and its $SL(2,\bb{R})$--related
partners, the R7- and T7-branes are new solutions which in the
present context occur on the same footing as the D7-brane. We also
find a G7 domain wall solution which has vanishing potential but
non-vanishing superpotential. It can be uplifted to a half--supersymmetic
conical space-time without scalars but with Killing
spinors, giving rise to a non-trivial monodromy.

In this Section we will present the explicit form of the solutions, both in D=9
and D=10, corresponding to the charges given in Table 2.

\subsection{Class I :\ $\alpha^2 = 0$}

We find the following half-supersymmetric domain wall solution
\begin{equation}
  \text{DW$_{\text{I}}$:~~~}\left\{\begin{aligned}
  ds^2 & = (C_1 m y)^{1/7}ds_8{}^2 + (C_1 m y)^{8/7}dy^2 \,,
\hskip 1.5truecm\\
  e^\phi & = (m y)^{-1}\,, \qquad
  e^\varphi = (C_1 m y)^{-2/\sqrt{7}} \,, \qquad
  \chi=C_2 \,,
\end{aligned}\right.
\label{D7in9gen}
\end{equation}
where the constant $C_2$ is arbitrary while $C_1$ is strictly positive.
The range of $y$ is such
that $m y$ is strictly positive in order to have a well-behaved metric.
We have used here the freedom of making a
re-parameterization in the transverse direction in order to make
the solution fall into the general class of seven-branes
\eqref{7brane} after uplifting to 10 dimensions.
We can also solve for the Killing spinor giving
\begin{equation}
\epsilon=(m y)^{1/28}\epsilon_0\, ,
\end{equation}
where $\epsilon_0$ is a constant spinor satisfying\footnote{The
chirality is determined by our convention that we choose the
transverse vielbein to be positive.}
$(1-i\Gamma_{\underline{xy}})\epsilon_0=0$.

Uplifting the above domain wall solution to 10 dimensions yields
(with $x$ being the reduction direction)
\begin{equation}
\text{D7:~~~~~}\left\{\begin{aligned}
\hat{d}s^2 & = ds_8^2+C_1 m y\left(dx^2+dy^2 \right)\,,\\
e^{\hat\phi} & = (m y)^{-1} \,,\\
\hat\chi & = m x + C_2 \, .\end{aligned}\right.
\end{equation}
 The uplifted Killing spinor is constant and still satisfies
$(1-i\Gamma_{\underline{xy}})\hat\epsilon=0$. The scalars and
spinors satisfy the monodromy requirement \eqref{monodromy} with
$a=1 ,\ b= 2\pi mR ,\ c=0 $ and $d=1$. We find that this solution
is a special case of the general seven-brane solution
\eqref{7brane} with
\begin{equation}
  f = m z+C_2\, ,\hskip 2truecm
  g = -\log(C_1) \, ,
\end{equation}
We can thus identify the two free parameters $C_1$ and
$C_2$ in the solution as coming from scalings (while keeping $m z$ fixed) and
shifts of the coordinates respectively.

\subsection{Class II :\ $\alpha^2 > 0$}

In this class we find the following half-supersymmetric domain wall
solution
\begin{equation}
  \text{DW$_{\text{II}}$:~~~}\left\{\begin{aligned}
  ds^2 & = \left(C_1 \cos{(m y)}\right)^{1/7}ds_8{}^2
    + \left(C_1 \cos(m y)\right)^{8/7}dy^2 \,, \hskip 1.5truecm\\
   e^\phi & = \left(e^{C_2}\cos{(m y)}\right)^{-1}\, , \qquad
   \chi = -e^{C_2} \sin(m y) \,, \\
   e^\varphi & = \left(C_1 \cos(m y)\right)^{-2/\sqrt{7}} \,,
   \end{aligned}\right.\label{DWII}
\end{equation}
where $C_2$ is arbitrary, $C_1$ is strictly positive and the range of $y$
has to be restricted so that $\cos{(m y)}$ is strictly positive.
The Killing spinor corresponding to the present solution is given by
\begin{equation}
  \epsilon=\left(C_1 \cos{(m y)}\right)^{1/28}
  e^{i m y/4}\epsilon_0 \,,
\end{equation}
where $(1-i\Gamma_{\underline{xy}})\epsilon_0=0$.

Note that in this class there is no solution with constant axion. This is
consistent with the fact that for zero axion the potential
corresponding to class II reads
\begin{equation}
 V(\varphi) = \tfrac{1}{2}m^2 e^{4 \varphi / \sqrt{7}}\, ,
\end{equation}
which, using the terminology of \cite{Lu:1996hm,Lu:1995cs}, is a
$\Delta = 0$ potential for which the standard domain wall solution
does not work.

The uplifting of this solution to 10 dimensions is given by
\begin{equation}
  \text{R7:~~~~~}\left\{\begin{aligned}
  \hat{d}s^2 & = ds_8^2+C_1 \cos(my) \left(dx^2+dy^2
  \right)\,,\\
  e^{\hat\phi} & = e^{-m x-C_2} \left(\cos( m y) \right)^{-1}\,,\\
  \hat\chi & = - e^{m x+C_2} \sin(m y)\,,\end{aligned}\right.
\end{equation}
where the Killing spinor is now given by
\begin{equation}
\hat\epsilon=e^{i m y/4}\hat\epsilon_0\,.
\end{equation}
The scalars and spinors
satisfy the monodromy requirement \eqref{monodromy} with
$a=e^{m\pi R} ,\ b= 0 ,\ c=0 $ and $d= e^{-m\pi R}$.
This solution falls in our general class of seven-branes
\eqref{7brane} with
\begin{equation}
  f = i e^{m z+C_2} \,, \qquad
  g = m z +C_2 - \log(C_1) \,.
\end{equation}
Thus the constants $C_1$ and $C_2$ have the same origin as in the
Class I solution: scalings and shifts of the coordinates.

\subsection{Class III :\ $\alpha^2 <0$}

In this class we have to divide into two subclasses depending on whether the
dilaton $\phi$ is non-zero (Class IIIa) or zero (Class IIIb).

\subsubsection{Class IIIa :\ $\alpha^2 <0$ and $\phi \neq 0$}

For non-zero dilaton we find the following half-supersymmetric
domain wall solution
\begin{equation}
  \text{DW$_{\text{IIIa}}$:~~~}\left\{\begin{aligned}
  ds^2 & =  \left(C_1 \sinh{(my)}\right)^{1/7}ds_8{}^2 +
     \left(C_1 \sinh{(my)}\right)^{8/7}dy^2 \,, \hskip 1.5truecm\\
  e^\phi & = {\cos(C_2)+\cosh{(m y)}\over \sinh{(m y)}}\,,\qquad
  \chi = {\sin(C_2) \over \cos(C_2) +\cosh{(m y)}}\,, \\
  e^\varphi & = \left(C_1 \sinh{(m y)}\right)^{-2/\sqrt 7} \,,
\end{aligned}\right.
\label{DWIIIa}
\end{equation}
where $C_2$ is an arbitrary angle between\footnote{It is of course
possible to extend the domain of $C_2$, but with the choice of
$-\pi/2$ to $\pi/2$ no solutions are related via $SL(2,\bbs{Z})$
and in this sense $C_2$ covers the space of solutions exactly
once.} $-\pi/2$ and $\pi/2$, $C_1$ is a strictly positive constant
and the range of $y$ is restricted by the requiring $\sinh(m y)$
to be strictly positive. The Killing spinor for this solution is
given by
\begin{equation}
  \epsilon=\left(C_1 \sinh(my)\right)^{1/28}e^{i\beta}\epsilon_0\, , \qquad
  \beta = \tfrac{1}{4} \text{arccot}\left(1+\cos(C_2) \cosh(my)
    \over \sin(C_2) \sinh(my)\right) \, ,
\end{equation}
where $(1-i\Gamma_{\underline{xy}})\epsilon_0=0$.

Lifting the solution \eqref{DWIIIa} to 10 dimensions gives
\begin{equation}
\text{T7:~~~~~}\left\{\begin{aligned}
\hat{d}s^2 & = ds_8^2 + C_1 \sinh(m y) \left(dx^2+ dy^2\right)\,,\\
e^{\hat\phi} & = {\cos(m x+C_2)+\cosh(m y)\over\sinh(m y)}\,,\label{classIIID10gen}\\
\hat\chi & = {\sin(m x+C_2)\over\cos(m x + C_2)+\cosh(m y)}\,.
\end{aligned}\right.
\end{equation}
The Killing spinor can also be lifted using (\ref{redf}) yielding
\begin{equation}
  \hat \epsilon = e^{i \beta}  \hat \epsilon_0 \,, \qquad
  \beta = \tfrac{1}{2} \text{arctan} \left( \text{tan}(\tfrac{1}{2}(mx+C_2)) \,
  \text{tan}(\tfrac{1}{2}my) \right) \,.
\end{equation}
Note that here the Killing spinor acquires non-trivial $x$--dependence.
We have explicitly checked that the monodromy requirement \eqref{monodromy}
is satisfied with $a = \cos(m\pi R),\ b= \sin(m\pi R),\ c = -\sin(m\pi R)$
and $d = \cos(m\pi R)$.
We note that also this class falls into the general class of
seven-brane solutions \eqref{7brane} with
\begin{align}
  f = \tan\left(\tfrac{1}{2}(m z +C_2)\right)\,, \qquad
  g = -2\log\left(\cos(\tfrac{1}{2}(mz+C_2))\right)-\log(C_1) \,.
\end{align}

\subsubsection{Class IIIb :\ $\alpha^2 <0$ and $\phi=0$}

For the case with vanishing dilaton we find the following
half--supersymmetric domain wall solution\footnote{This solution
  is related by a coordinate transformation to that of \cite{Cowdall:2000sq},
  where, contrary to our result, it was claimed that in order to preserve a fraction of
  the supersymmetry $m$ should be zero, reducing the solution to Minkowski
space-time
  with arbitrary constant scalars.}:
\begin{equation}
  \text{DW$_{\text{IIIb}}$:~~~}\left\{\begin{aligned}
  ds^2 & = e^{my/7} ds_8{}^2 + e^{8my/7} dy^2 \,, \\
  e^\varphi & = e^{-2my/\sqrt{7}} \,, \qquad \phi = \chi = 0 \,,
\end{aligned}\right.
\label{DWIIIb}
\end{equation}
where the range of $y$ is unrestricted. The corresponding Killing spinor reads
\begin{align}
  \epsilon = e^{my/28} \epsilon_0 \,,
\end{align}
with $(1-i\Gamma_{\underline{xy}})\epsilon_0=0$.

We note that for this solution, since $\chi = 0$, the potential and
superpotential read
\begin{equation}
 V(\phi,\varphi) = \tfrac{1}{2}m^2 e^{4 \varphi / \sqrt{7}}
    \sinh^2(\phi)\, ,\qquad
 W(\phi,\varphi) =\tfrac{1}{4}m\, e^{2\varphi/\sqrt{7}} {\rm cosh}\, \phi\, .
\end{equation}
The above potential has occurred recently, see eq.~(77) of
\cite{Kallosh:2001gr}, in the context of a possible inflation
along flat directions. An interesting feature of this case is that
the flat direction, $\phi=0$, corresponds to a vanishing
potential, $V(\varphi)=0$, despite a non--vanishing superpotential, $W
= \tfrac{1}{4}m \, e^{2\varphi/{\sqrt{7}}}$.
Such a situation has occurred recently in the context of
quintessence in N=1 supergravity, see Section 3 of
\cite{Townsend:2001ea}.

Lifting this solution to 10 dimensions leads to
the following purely gravitational solution\footnote{Other examples
of domain walls that lift up to purely gravitational solutions
have been given in \cite{Gibbons:2001ds}.}:
\begin{equation}
\text{G7:~~~~~}\left\{\begin{aligned}
  \hat{d}s^2 & = ds_8^2 + e^{m y} \left(dx^2+ dy^2 \right)\,,\\
  {\hat\phi} & = \hat\chi = 0\, .
\end{aligned}\right.
\end{equation}
For the lifted Killing spinor we find
\begin{equation}
  \hat\epsilon=e^{i m x /4}\epsilon_0\, .
\end{equation}
Again, this solution falls in the class of purely gravitational
solutions discussed in Section 3.1 with the identifications
\begin{align}
  f =i \,, \qquad g = i m z \,.
\end{align}
As discussed in section 3.2, the holomorphic function $g$ can be
transformed away: the coordinate transformation \eqref{holomorphicgct} takes
the form $r=\frac{2}{m}e^{m y/2}$ and $\theta=\tfrac{1}{2}(\pi- mx)$. The
compactness of $x$ translates into $\theta \sim \theta + m \pi R$. 

We can now impose three different quantization conditions (to be discussed in Section 5).
The condition $m=1/(2 R)$ implies that this solution describes a conical space-time
with deficit angle $3\pi/2$. In other words, this is a half--supersymmetric 
$\text{Mink}_8 \times \mathbb{C} / \mathbb{Z}_4$ space-time with non-trivial monodromy, 
the bosonic part of which was also mentioned in \cite{Greene:1990ya}. The second
quantization condition $\tilde m = 1/(3 \sqrt{3} R)$ can only be applied to an 
$SL(2,\mathbb{R})$--related partner of the G7-brane and gives rise to a deficit
angle of $5\pi/3$. This is a half--supersymmetric 
$\text{Mink}_8 \times \mathbb{C} / \mathbb{Z}_6$ space-time with non-trivial monodromy.
The third quantization condition $m=2/R$ yields the identification 
$\theta \sim \theta + 2 \pi$ and indeed this is fully supersymmetric $\text{Mink}_{10}$ space-time. 
The monodromy is trivial and there is a second Killing spinor $\hat\epsilon=e^{-i m x /4}\epsilon_0$ 
with opposite chirality. For the previous two quantization conditions this second
Killing spinor had a different monodromy and was therefore not consistent.

\subsection{Class IV : $m_4$}

We first substitute the domain wall Ansatz in the Killing spinor
equations, which are in this class given by (\ref{class4susy}). We
find that we can not construct a projector, yielding a 1/2
supersymmetric domain wall, out of the $\Gamma$-matrices appearing
in the supersymmetry rules unless the scalars have a
time-dependence, i.e. the transverse direction has to be the time
direction. In this respect class IV is fundamentally different from
class I-III. For time-dependent solutions we cannot assume that a
solution to the Killing spinor equations is automatically a
solution to the full equations of motion. A counter-example is
provided by considering a scalar $\Phi$ that does only occur in
the transformations of the spin-1/2 fermions as $(\not \!\! \partial
\Phi)\epsilon$. Clearly the Killing spinor equations can be solved
for a flat metric and $\Phi = \Phi(u), \gamma_v\epsilon = 0$ where
we use lightcone coordinates $u = x+t, v = x-t$. The non-zero
scalar leads to a nonzero $uu$-component of the energy-momentum
tensor and the Einstein equations are not solved.

On the other hand, examples of time--dependent 1/2 supersymmetric
BPS solutions are known. An example is the gravitational wave
solution. For the present case, however, we find that it is not
possible to construct a domain wall solution, time-dependent or
not, preserving any fraction of the supersymmetry.

\section{Quantization Conditions}

It is well-known that at the quantum level
the classical $SL(2,\bb{R})$ symmetry of IIB supergravity is broken
to $SL(2,\bb{Z})$\footnote{A similar quantization condition
does not apply to the $SO(1,1)$ symmetry of IIA supergravity.}.
We would like to consider the
effect of this on the solutions discussed in the previous sections.
In particular, it implies that the monodromy matrix must be an element of
the arithmetic subgroup of $SL(2,\bb{R})$:
\begin{align}
  \hat{\mathcal M} (x+2\pi R) = \Lambda \, \hat{\mathcal{M}}(x) \Lambda^T
  \qquad \text{with~~} \Lambda = e^{2 \pi R \, C} \in SL(2,\bb{Z}) \,.
\label{quantcond}
\end{align}
This will imply a charge quantisation of the 7--brane solutions in 10D.
Since these charges give rise to the mass parameters upon reduction,
at the same time this requirement therefore implies a mass quantisation.

We will apply the following procedure. The mass parameters will be
parameterized by $\vec{m}=\tilde m \, (p,q,r)$. Then, given the radius of
compactification $R$ and the relative coefficients $(p,q,r)$ of the mass
parameters, one should choose the overall coefficient $\tilde m$
such that the monodromy lies in $SL(2,\bb{Z})$. This is not always possible;
a necessary requirement in all but one cases will be that $(p,q,r)$ are
integers and satisfy a diophantic equation. Furthermore we must
require $q$ and $r$ to be either both even or both odd. Only in Class III
will it be possible to quantize for non-integer $(p,q,r)$. Thus we get all
$SL(2,\bb{Z})$ monodromies that can be expressed as products of $S$ and $T$
(and their inverses) as defined in \eqref{SandT}.
These $SL(2,\bb{Z})$ conjugacy classes have been classified
in \cite{DeWolfe:1998eu, DeWolfe:1998pr}. The ones corresponding
to class I and class III have also been discussed in \cite{Hull:2002wg}.
The situation is summarized in Table 3.
Below the Table we consider each of the three classes separately.

\begin{table}[ht]
\begin{center}
\begin{tabular}{||c|c|c|c||c|c||c||}
  \hline \rule[-1mm]{0mm}{6mm}
  Class & $\alpha^2$ & $\text{Tr}(\Lambda)$ & $p^2+q^2-r^2$ &
    $(p,q,r)$ & $\Lambda$ & \\
  \hline \hline \rule[-5mm]{0mm}{12mm}
  I & $=0$ & $2$ & $0$ & $(0,n,n)$
    & $T^n= \left( \begin{array}{cc} 1 & n \\ 0 & 1 \end{array}\right)$
    & $n\in \bb{Z}$ \\
  \hline \rule[-5mm]{0mm}{13mm}
  II & $>0$ & $n$ & $n^2-4$ & $(\pm n,0,\pm 2)$
     & $(S \, T^{-n})^{\pm 1}=
     \left( \begin{array}{cc} 0 & 1 \\ -1 & n \end{array} \right)^{\pm 1}$
     & $3 \leq n\in \bb{Z}$ \\
  \hline \rule[-5mm]{0mm}{13mm}
  III & $<0$ & $0$ & $-4$ & $(0,0,\pm 2)$
      & $S^{\pm 1}=
      \left( \begin{array}{cc} 0 & 1 \\ -1 & 0 \end{array} \right)^{\pm 1}$
      & \\
    & & $1$ & $-3$ & $(\pm 1,0,\pm 2)$ & $(T^{-1} \, S)^{\pm 1}=
      \left( \begin{array}{cc} 1 & 1 \\ -1 & 0 \end{array} \right)^{\pm 1}$
      & \\ \rule[-5mm]{0mm}{12mm}
    & & $2$ & $-4$ &
      & $\id=\left( \begin{array}{cc} 1 & 0 \\ 0 & 1 \end{array} \right)$
      & \\
\hline
\end{tabular}
\caption{\it The table summarizes the different
$SL(2,\bb{Z})$--monodromies. It is organized according to the trace of
the monodromy and gives the diophantic equation for $(p,q,r)$. Explicit
examples are given with the corresponding monodromies. For Case I and III
all diophantic solutions are related by $SL(2,\bb{Z})$ to the examples given.
In Case II there are other conjugacy classes \cite{DeWolfe:1998eu, DeWolfe:1998pr}.}
\end{center}
\end{table}
\begin{itemize}
\item
For class I with $\alpha^2=0$ the monodromy matrix reads
\begin{equation}
  \Lambda = \left(
    \begin{array}{cc} 1 + m_1 \pi R &
     \ \ \ (m_2+m_3) \pi R \\
        &\\
     (m_2-m_3) \pi R &
     \ \ \ 1 - m_1 \pi R \end{array}
    \right) \,.\label{monI}
\end{equation}
We find that $\Lambda$  is an element of $SL(2,\bb{Z})$ provided we have
\begin{align}
  \text{Class I:~~} \tilde m = \frac{1}{2\pi R}
  \text{~~and~~} p^2+q^2-r^2=0 \,.
\end{align}
All the solutions of the diophantic equation are related
via $SL(2,\bb{Z})$ to the D7-brane solutions with $(p,q,r)=(0,n,n)$
with $n$ an arbitrary integer
\cite{DeWolfe:1998eu, DeWolfe:1998pr,Hull:2002wg},
which is the explicit choice we
have used for Class I. This gives rise to the monodromy
$\Lambda=T^n$. The quantization on $\tilde m$ is the same charge
quantisation condition as found in \cite{Bergshoeff:1996ui}.

\item
For class II with $\alpha^2>0$ the monodromy matrix reads
\begin{equation}
  \Lambda  = \left(
    \begin{array}{cc} \cosh(\alpha 2 \pi R) + \tfrac{m_1}{2\alpha} \sinh(\alpha 2 \pi R) &
     \ \ \ \tfrac{1}{2 \alpha} (m_2+m_3) \sinh(\alpha 2 \pi R) \\
        &\\
     \tfrac{1}{2 \alpha} (m_2-m_3) \sinh(\alpha 2 \pi R) &
     \ \ \ \cosh(\alpha 2 \pi R) - \tfrac{m_1}{2\alpha} \sinh(\alpha 2 \pi R) \end{array}
    \right) \, .
\end{equation}
We find that $\Lambda$  is an element of $SL(2,\bb{Z})$ provided we have
\begin{align}
  \text{Class II:~~} \tilde m = \frac{\text{arccosh}(n/2)}{\pi R\sqrt{n^2-4}}
  \text{~~and~~} p^2+q^2-r^2=n^2-4 \,,
\end{align}
for some integer $n \geq 3$. This has solutions $(p,q,r)=(\pm n,0,\pm 2)$ with
monodromy $\Lambda=(S \, T^{-n})^{\pm 1}$ but not all other solutions are
related by
$SL(2,\bb{Z})$ \cite{DeWolfe:1998eu, DeWolfe:1998pr}. Note that the explicit
choice we have made for Class II with $(p,q,r)=(p,0,0)$ does not solve
the diophantic equation. Thus the R7-brane is not consistent at the
quantum level but particular $SL(2,\bb{R})$ partners are.

\item
For class III with $\alpha^2<0$ the monodromy matrix reads
(using $\alpha =i m$)
\begin{equation}
  \Lambda  = \left(
    \begin{array}{cc} \cos(m 2 \pi R) + \tfrac{m_1}{2m}
       \sin(m 2 \pi R) &
     \ \ \ \tfrac{1}{2 m} (m_2+m_3) \sin(m 2 \pi R) \\
        &\\
     \tfrac{1}{2 m} (m_2-m_3) \sin(m 2 \pi R) &
     \ \ \ \cos(m 2 \pi R) - \tfrac{m_1}{2m}
       \sin(m 2 \pi R) \end{array}
    \right) \,.
\end{equation}
Here we find that there are three distinct possibilities for $\Lambda$ to
be an element of $SL(2,\bb{Z})$. For the first possibility we must have
\begin{align}
  \text{Class III:~~} \tilde m = \frac{1}{4 R}
  \text{~~and~~} p^2+q^2-r^2=-4 \,.
\end{align}
This is the explicit choice we have made for the T7- and G7-brane solution
with $(p,q,r)=(0,0,\pm2)$ and $\Lambda=S^{\pm 1}$. In fact all other
solutions to the diophantic equation are related by $SL(2,\bb{Z})$
\cite{DeWolfe:1998eu, DeWolfe:1998pr,Hull:2002wg}. For the second possibility one must require
\begin{align}
  \text{Class III:~~} \tilde m = \frac{1}{3 \sqrt{3} R}
  \text{~~and~~} p^2+q^2-r^2=-3 \,.
\end{align}
We have not explicitly considered this case but one solution is
$(p,q,r)=(\pm1,0,\pm2)$ with monodromy $\Lambda=(T^{-1} \, S)^{\pm1}$.
Again all other solutions are related by $SL(2,\bb{Z})$
\cite{DeWolfe:1998eu, DeWolfe:1998pr,Hull:2002wg}.
If neither of these two possibilities applies one can always choose
\begin{align}
  \text{Class III:~~} \tilde m = \frac{1}{R}
  \text{~~and~~} p^2+q^2-r^2=-4 \,,  \; \; (p,q,r) \in \bb{R} \,,
\end{align}
where $(p,q,r)$ are not required to be integer-valued. This gives rise
to trivial monodromy $\Lambda=\id$.
\end{itemize}

\section{Conclusions}

One of the aims of this letter is to study domain wall solutions in a
9--dimensional setting. The advantage of picking out 9 dimensions is that
it is simple enough to investigate in full detail but also shares some of
the complications of the lower--dimensional supergravities. In Section 4
we constructed several half--supersymmetric domain wall solutions and we
gave their uplifting to 10 dimensions. This uplifting introduces a
nontrivial dependence of the D=10 solution on the compact coordinate.

In the IIB case we thus found three classes of 7--brane solutions in 10
dimensions, which are all characterized by two holomorpic functions
\eqref{7brane}. One class contains the D7-brane and its $SL(2,\bb{R})$--related
partners but the R7--brane and T7--brane solutions we found for the
other two classes are not related to the
D7--brane by $SL(2,\bb{R})$--duality. It would be interesting to
see what their interpretation is terms of the type IIB superstring theory.
Together the solutions provide a set of half-supersymmetric
7--branes with arbitrary charges that are consistent in the sense that
the monodromies of the scalars and Killing spinors coincide.
Our method of uplifting domain walls also leads to half--supersymmetric 
conical G7--brane solutions
with deficit angles $3\pi/2$ or $5\pi/3$, not carried by any scalars.
The nontrivial monodromies sit in the fermionic sector. The G7--brane 
solution can also be uplifted to Minkowski space-time, in which case we
have supersymmetry enhancement upon uplifting.
It would be interesting to further study the properties of
the D=9 domain wall solutions and their D=10 7--brane origins and to see
whether some of the features we find also occur for $D<9$ domain walls.

The three distinct massive supergravities corresponding to the IIB case are
$SL(2,\bb{R})$ covariant and characterized by the $SL(2,\bb{R})$
invariant $\alpha^2$. One of them, with $\alpha^2 = 0$, has a singular
mass matrix and therefore, following a similar statement made in
\cite{Alonso-Alberca:2000gh}, does not seem to correspond to a gauged
supergravity theory.
The class with $\alpha^2 < 0$ has been shown to be an
$SO(2)$-gauged supergravity \cite{Cowdall:2000sq}. We conjecture
that the remaining class with $\alpha^2 >0$ is an $SO(1,1)$-gauged
supergravity. Interestingly, in a recent paper it is stated that both
the $\alpha^2 = 0$ and the $\alpha^2>0$ cases correspond to
$SO(1,1)$-gauged supergravities \cite{Hull:2002wg}.
The distinction between the different theories does not occur in the
compact case,
i.e.~when the symmetry group would be $SU(2)$ rather than $SL(2,\bb{R})$.
Such a situation occurs for instance when gauging the $U(1) \subset SU(2)$
R--symmetry group in N=2, D=5 supergravity coupled to vector
multiplets. Here all choices for the mass parameters are physically
equivalent leading to a single gauged supergravity theory,
see e.g.~\cite{Bergshoeff:2000zn}.

In the IIA case we performed two reductions, one leading to the
$m_4$--deformation and one leading to class I. In the case of the
$m_4$--deformation we find that there is no domain wall solution
preserving any supersymmetry.
The reason that we could not
perform both IIA reductions at the same time was that the $SO(1,1)$--symmetry
is only valid for $m_{\rm R}=0$. One might change this situation
by replacing $m_{\rm R}$ by a scalar field $M(x)$ and a 9--form
Lagrange multiplier $A^{(9)}$ \cite{Bergshoeff:2001pv} via:
\begin{equation}
  \mathcal{L}(m_{\rm R}) \rightarrow \mathcal{L}(M(x)) + M(x)\partial A^{(9)}\, .
\end{equation}
Unfortunately, the reduction of the second term leads to an
additional term in 9 dimensions containing a 9--form Lagrange multiplier.
The equation of motion of this Lagrange multiplier leads to the
constraint $M(x)m_4 = 0$ which brings us back to the previous situation.
A similar thing happens in 11 dimensions if one tries to use
the same trick to convert the scale symmetry of the
equations of motion to a symmetry of the action by replacing the
gravitational constant by a scalar field. The elimination of the
Lagrange multiplier brings us back to the analysis of \cite{Howe:1998qt}.

Let us finally comment on the relation between the
different massive deformations of N=2 D=9
supergravity and T-duality. The massless theory can be obtained
from the reduction of both IIA and IIB massless supergravity
\cite{Bergshoeff:1995as}. This follows from the T--duality between the
underlying IIA and IIB string theories. However, the Scherk--Schwarz
reductions
of IIA and IIB supergravity to nine dimensions give rise to four different
massive
deformations of the unique massless theory.
Only one of these deformations (Class I)
can be reproduced by both IIA and IIB supergravity.
It is not clear what the IIA or M-theory origin is of the other
two deformations (class II and class III)\footnote{We do not consider here
the use of Killing vectors in the Lagrangian. Assuming that the IIA theory
has such explicit Killing vectors a massive T-duality map can be constructed
\cite{Meessen:1998qm}. We neither consider a further reduction to
D=8 dimensions \cite{Hull:1998vy}.}. Similarly, it is not clear what the
IIB origin is of the class IV deformation.

To understand massive T-duality it might be necessary to explicitly
include massive winding multiplets\footnote{The inclusion of the full
tower of (massive) multiplets of higher Fourier and winding number has
been discussed in
\cite{Abou-Zeid:1999fv,deWit:2000wu,deWit:2000zz}.}
(while in supergravity reduction
one only keeps the states without winding).
Massive T-duality suggests the existence of a maximally
supersymmetric
massive supergravity theory containing {\it all four} mass parameters
$(m_1, m_2, m_3, m_4)$. The existence of such a theory is not implied by the
massive supergravities with seperate deformations $(m_1,m_2,m_3)$ and $m_4$.
This massive supergravity has already been
suggested for
different reasons in \cite{Meessen:1998qm} and it would be interesting to
see whether it can be constructed \cite{Bergshoeff:2002}.

\section*{Acknowledgments}

We thank Jisk Attema, Gabriele Ferretti, Chris Hull, Roman Linares, Bengt E.W. Nilsson, 
Tom\'as Ort\'\i n
and Tim de Wit for useful discussions. D.R. would also like to
thank Per Sundell for interesting and useful discussions in the
early stages of the work. E.B. would like to thank the Newton Institute
for Mathematical Sciences in Cambridge, where part of this
work was done, for hospitality. This work is supported in part by the
European Community's Human Potential Programme under contract
HPRN-CT-2000-00131 Quantum Spacetime, in which E.B., U.G.~and
D.R.~are associated to Utrecht University. The work of U.G.~is
part of the research program of the ``Stichting voor Fundamenteel
Onderzoek der Materie'' (FOM).

\appendix
\section{Conventions}
We use mostly plus signature $(-+\cdots +)$. Hatted fields
and indices are ten-dimensional while unhatted ones
are nine-dimensional. Greek indices
$\hat \mu,\hat \nu,\hat \rho\ldots$ denote world coordinates and Latin indices
$\hat a,\hat b,\hat c\ldots$ represent tangent spacetime. They are related by the
vielbeins $\hat e_{\hat \mu}{}^{\hat a}$ and inverse vielbeins $\hat e_{\hat a}{}^{\hat \mu}$.
Explicit indices $x,y$ are underlined when flat and non-underlined when curved.
We antisymmetrize with weight one, for instance
$(\partial \hat A)_{\hat \mu \hat \nu} =
 \tfrac{1}{2} (\partial_{\hat \mu} \hat A_{\hat \nu} - \partial_{\hat \nu} \hat A_{\hat \mu})$.
Omitted indices are contracted without numerical factors,
e.g.~$(\partial \hat A)^2 = (\partial \hat A)_{\hat \mu \hat \nu}
(\partial \hat A)^{\hat \mu \hat \nu}$. The covariant derivative
on fermions is given by $D_{\hat{\mu}}=\partial_{\hat{\mu}}+
\hat{\omega}_{\hat \mu}$ with the spin connection
$\hat{\omega}_{\hat \mu} = {\textstyle{1\over 4}}
\hat{\omega}_{\hat{\mu}}{}^{{\hat{a}} {\hat{b}}}
{\Gamma}_{{\hat{a}}{\hat{b}}}$.

We have chosen all $\Gamma$-matrices real. Curved indices of hatted $\Gamma$-matrices $\hat \Gamma_{\mu}$
refer to the ten-dimensional metric while curved indices of unhatted $\Gamma$-matrices $\Gamma_{\mu}$
refer to the nine-dimensional metric. Furthermore
\begin{align}
  \Gamma_{11} = \Gamma^{\underline{0} \cdots \underline{9}} \,, \qquad \Gamma_{11}{}^2 = 1 \,.
\end{align}
In the IIA theory we have real Majorana spinors of indefinite chirality. In the IIB theory we
have complex spinors of definite chirality. To switch between Majorana and Weyl fermions in nine
dimensions one must use
\begin{equation}
\begin{aligned}
  \tfrac{1}{2} (1+\Gamma_{11}) \psi_\mu^M & = \text{Re} (\psi_\mu^W)\, , \\
  \tfrac{1}{2} (1+\Gamma_{11}) \lambda^M & = \text{Im} (\Gamma_{\underline{x}} \lambda^W)\,, \\
  \tfrac{1}{2} (1+\Gamma_{11}) \tilde\lambda^M & = \text{Im} (\Gamma_{\underline{x}} \tilde\lambda^W)\,, \\
  \tfrac{1}{2} (1+\Gamma_{11}) \epsilon^M & = \text{Re} (\epsilon^W)\,,\qquad\qquad
\end{aligned}
\begin{aligned}
  \tfrac{1}{2} (1-\Gamma_{11}) \psi_\mu^M & = \text{Im} (\Gamma_{\underline{x}} \psi_\mu^W)\, , \\ 
  \tfrac{1}{2} (1-\Gamma_{11}) \lambda^M & = \text{Re} (\lambda^W)\, , \\
  \tfrac{1}{2} (1-\Gamma_{11}) \tilde\lambda^M & = \text{Re}
    (\tilde\lambda^W)\, , \\
  \tfrac{1}{2} (1-\Gamma_{11}) \epsilon^M & = \text{Im} (\Gamma_{\underline{x}} \epsilon^W)\, ,
\end{aligned}
\end{equation}
for positive ($\psi_\mu^W, \epsilon^W$) and negative ($\lambda^W,\tilde\lambda^W$) chirality Weyl fermions.

\providecommand{\href}[2]{#2}\begingroup\raggedright\endgroup

\end{document}